\documentclass[12pt]{article}
\usepackage{epsfig,times}
\usepackage{picinpar}
\usepackage{wrapfig}
\usepackage{floatflt}
\newfont{\preprint}{lasy10 scaled 1500}

\newcommand{\AmS}{{\protect\the\textfont2
 A\kern-.1667em\lower.5ex\hbox{M}\kern-.125emS}}
\makeatletter
\renewcommand{\section}{\@startsection{section}{1}{0in}
    {0.4\baselineskip}{0.1\baselineskip}{\large\bf}}
\renewcommand{\subsection}{\@startsection{subsection}{2}{0in}
    {0.25\baselineskip}{-\baselineskip}{\large\bf}}
\renewcommand{\subsubsection}{\@startsection{subsubsection}{3}{0in}
    {0.1\baselineskip}{-\baselineskip}{\normalsize\bf}}
\makeatother
\begin{document}
{\large
\begin{center}
     On the Velocity of Light Signals in the  Deep
     Underwater Neutrino Experiments
\end{center}
}
%\vspace{0.5cm}
\begin{center}
               L.A.Kuzmichev\\
    Skobeltsyn Institute of Nuclear Physics, Moscow State University\\
   119899, Moscow,Russia, e-mail: kuz@dec1.npi.msu.su
\end{center}
\begin{abstract}
    During the last few years  deep underwater neutrino telescopes
    of a new generation with dimensions close to 100 m or more were
    taken into operation. For the correct track reconstruction of events
    and for the interpretation of light pulses from calibration lasers
    one has to use the
    group  velocity for light signals.
    The difference between group velocity and the traditionally used
    phase velocity leads to an additional delay of about
    10 ns for a distance of 100 m between light source and
    photomultiplier.
    From the time of the appearance of the first projects of
    deep underwater neutrino telescopes in the middle of 70th
    this fact was never mentioned in the literature.
\end{abstract}
\section*{}
   During the last few years the new  large - scale neutrino telescopes
   NT-200 /1/ and \\
   \mbox{AMANDA /2/} have been taken into operation. Two
   deep underwater arrays, \mbox{NESTOR/3/} and ANTARES /4/, are under
   construction.

   The Cherenkov light from high energy muons, electromagnetic and
   hadronic showers can be recorded at distances of 20 -- 100 m 
   depending
   upon the light absorption of water or ice.  Impulse laser light sources
   are widely  used in these arrays for calibration.
   Light pulses from such sources can be recorded over
   even larger distances. 

   At such distances, the difference between group and phase velocity
   of light in water or ice is essential.
   As  far as I know, this fact hasn't been mentioned in the literature from the
   time of the first neutrino telescope projects in the middle of
   the seventies 
   /5,6/ and hasn't been taken into account for the data analysis and
   MC calculations.

    The velocity of light pulses in matter is given by the
    group velocity /7,8/ :
$$
       V_{gr} = d\omega/dk \eqno (1)
$$
    where $\omega$ is the frequency, $k$ is the wave vector.

  The connection between $\omega$ and $k$ can be written in the following form
     ( dispersion equation):
$$
     \omega= c\cdot k/n(\omega) \eqno (2)
$$
     where $c$ is the velocity of light in vacuum, $n$ is the refraction index.

      As it is known, refraction indices given in handbooks like /9/
       correspond
      to the phase velocity due to the method they are measured /10/ :
$$
           n(\omega) = c/V_{ph}(\omega)  \eqno (3)
$$
            where $V_{ph}$ is the phase velocity.
\newpage
   It is reasonable to accept a group refraction index for the group velocity
       similar to the phase index of refraction:
$$
          n_{gr}(\omega) = c/V_{gr}(\omega)    \eqno (4)
$$
                where $V_{gr}$ is the group velocity.

    Using (2) and replacing the $\omega$ dependence of $n$ by a 
    wavelength dependence, the relation
    between  $n_{gr}$ and $n$ can be derived:

$$
      n_{gr}(\lambda) = n(\lambda) -\lambda dn/d\lambda  \eqno (5)
$$
                where  $\lambda$ is the wavelength.

\begin{figure}%[H]
\mbox{\epsfig{file=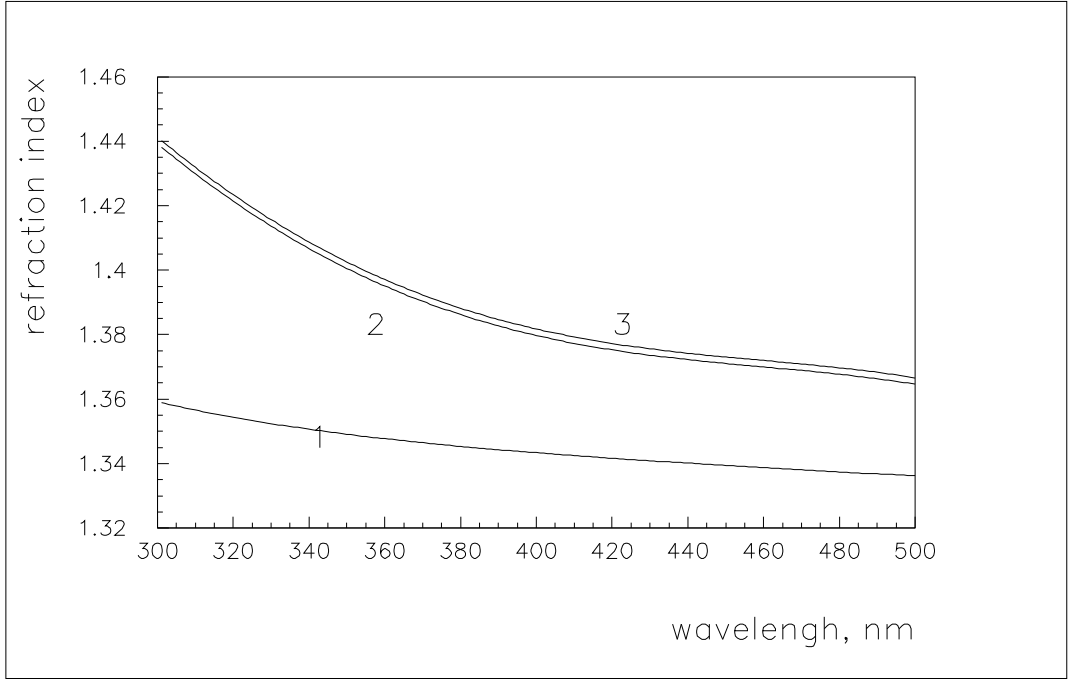,height=9.cm,width=14.0cm}}
\vspace{-3mm}           %\hspace{1.0cm}
%\parbox[b]{4.5cm}{\caption [2] {
{\caption [2] {
%\frenchspacing
The dependence of $n$ on wavelength (curve 1),$n_{gr}$ (curve 2) and $n_{gr}$ for
1 km depth (curve 3).
%\nonfrenchspacing
}}
\vspace{-3mm}
\end{figure}

     $dn/d\lambda$ is negative for water over the visible light range,
     and so $n_{gr}$ is larger than $n$.
     Fig.1 shows the wavelength dependence of $n$ ( curve 1) for
     destilled water at 20$^0$C /9/, and the same for $n_{gr}$ 
     (curve 2).

      The density of water influences the refraction
     index at large depths of neutrino telescopes.
     But this correction is
     rather small due to the comparatively small compression coefficient
     of water.
     For the conditions of NT-200 ( depth 1100 m,
     water temperature \mbox{4$^0$C)} the variation of the refraction index is
     about 0.002 only. This is more than 10 times smaller than the difference
     between $n_{gr}$ and $n$. This correction of $n$ would reach 0.01
     for the arrays to be placed in the ocean depths of 4 -5 km, but
     even in this case the depth correction is significantly smaller than
     that due to the difference between group and phase velocities.
      The curve 3 on fig.1 shows the dependence of $n_{gr}$ on $\lambda$ for
      the conditions of the telescope NT-200.

       All  curves at fig.1 were calculated for  destilled
      water. The light absorption of natural water differs from that
     of  destilled water, and so there is a difference in the image part
     of their dielectric penetration.
     As Kramers-Kroning equation connects the real part of the
     dielectric penetration with it's imaginary part /7/,  the real
     refraction index differs, in principle,  from that of destilled water.
     Such  a correction would be non - negligible only if there are the
     intensive lines of absorption in the visible light absorption spectrum.
     
\begin{figure}%[H]
\mbox{\epsfig{file=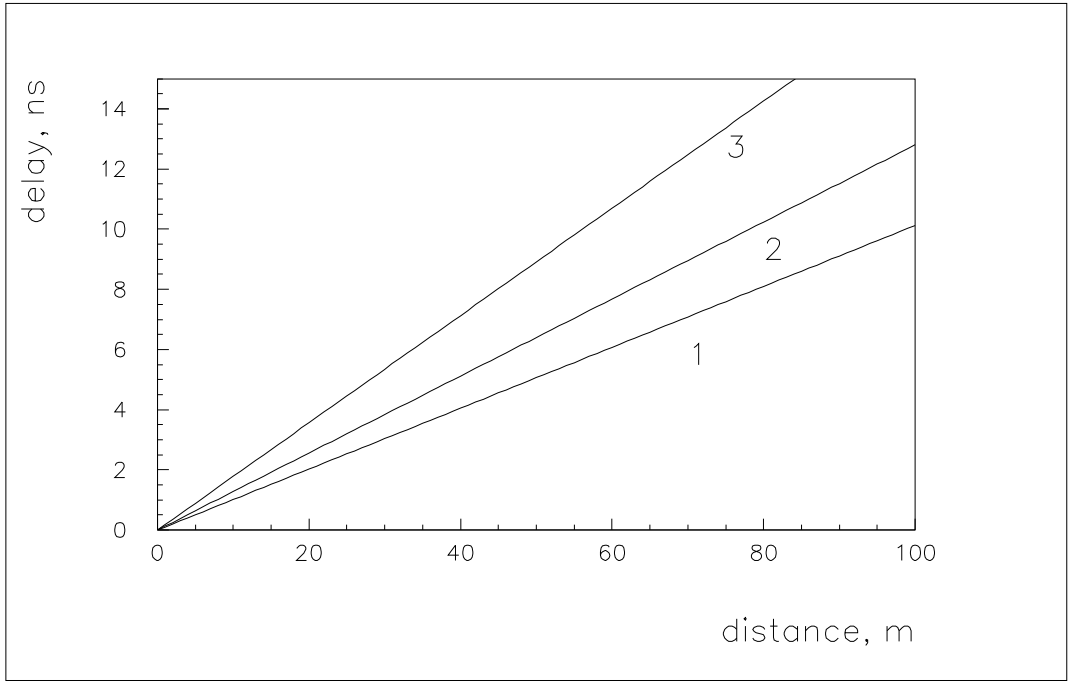,height=9.cm,width=14.0cm}}
\vspace{-3mm}           %\hspace{1.0cm}
%\parbox[b]{4.5cm}{\caption [2] {
{\caption [2] {
%\frenchspacing
The dependence of the delay on the distance, for wavelength 500 nm (curve 1),
400nm (curve 2) and 350nm (curve 3).
%\nonfrenchspacing
}}
\vspace{-3mm}
\end{figure}

     Fig.2 shows the dependencies of delays on the distance between light
     source and receiver which emerge from  the difference between the phase
     and group velocities for different wavelengths.

%      Use of the group velocity in calculations gives the possibility
%     of correct using of laser sources for investigation of
%    location of optical modules and for calibration of array
%     time systems, eliminates the discrepancies in time
%     distributions between experiment and MC calculation.

       For arrays with angular resolution 1-2$^0$ replacement of the
   phase velocity by the group velocity probably would not lead to 
   essential changes  in track reconstruction. For the projects which
   claim an angular accuracy 0.1-0.2$^0$ and 
   absorption  length of $\ge 50$m /4/ the use of
   group velocity in track reconstruction procedures seems to be absolutely
   necessary.

\section*{Acknowledgments}

    I would like to express thanks to G.V.Domogatsky, J-A.M. Djilkibaev,
    S.I.Klimushin, B.K.Lubsandorzhiev, E.A.Osipova, V.V.Prosin,
    C.Spiering and V.B.Tsvetkov for useful discussion.

\section*{ References}
\noindent
 1. I.A.Belolaptikov et al.,\\ 
 The Baikal Underwater Neutrino Telescope: Design, Performance, 
 and First Results.\\
 Astroparticle Physics 7(1997) 263-282.\\
 2.E.Andres et al.,\\
 The AMANDA Neutrino Telescope: Principle of Operation and First Results.\\
 Astroparticle Physics 13 (2000) 1-20.\\
\newpage
\noindent
 3.B.Monteleoni et al.,\\
 NESTOR a deep sea physics laboratory for the Mediterranean \\
 Proc. of the 17th International Conference on Neutrino Physics and Astrophysics,
 (Neutrino 96)\\
 Helsinki, Finland, June 13-19, 1996. \\
 4.Ph.Amram et al.,\\
 The ANTARES Project\\ 
 Nucl.Physics B (Proc.
 Suppl.) 75A (1999) 415.\\ 
 5.V.S.Berezinsky and G.T.Zatsepin\\
 Sov.Phys.Usp.
 20 (1977) 361. \\
 6.DUMAND-75.
 Proceeding of the 1975 Dumand Summer Study,Bellingham,\\ 
 edited by P.Kotzer.Published by Western Washington State College.\\
 7.L.D.Landau , E.M.Lifshiz.
 Electrodynamics of Continuous Media.\\ 
 Moscow, Nauka, 1982, p 345.\\
 8.A.Sommerfeld, OPTIK  , Wiesbaden 1950.\\ 
 9.Physical Quantities , Handbook,ed.
 I.S.Grigorev and E.Z.Meylihoba,\\
 Energoatomizdat, Moscow 1991, p.
 790. \\
 10. A.I.Stogarov and N.F.Timofeeva, \\
     Trudi GOI, 1963, v 31, N 160 (in Russian). 
\end{document}